# Double perovskite Ba$_2$NaOsO$_6$: Magnetic structure, Kramers doublet and x-ray magnetic dichroic signals

S. W. Lovesey

*ISIS Facility, STFC, Didcot, Oxfordshire OX11 0QX, UK*
*Diamond Light Source, Harwell Science and Innovation Campus, Didcot, Oxfordshire OX11 0DE, UK*
*Department of Physics, Oxford University, Oxford OX1 3PU, UK*

**Abstract** Of all the oxides of osmium, Ba$_2$NaOsO$_6$ is the only one with a substantial ferromagnetic moment. A fresh theoretical study of bulk magnetic properties of the double perovskite is accomplished with an appropriate space group, and an atomic model of a Kramers doublet formed by the single osmium d-electron. It possesses symmetry demanded by the Wyckoff position assigned to Os in a canted ferromagnet. Whereupon, magnetic moments in the paramagnetic and magnetically ordered states of Ba$_2$NaOsO$_6$ are unified. Analytic x-ray magnetic dichroic signals (XMCD) at osmium L$_2$ and L$_3$ absorption edges are presented, and shown to agree with hitherto unexplained recent measurements.

## I. INTRODUCTION

Among oxides of osmium, including binary OsO$_2$ and double and triple perovskites, Ba$_2$NaOsO$_6$ is the only one with a substantial ferromagnetic moment in the ordered state. It is substantially larger than in iridates, e.g., Sr$_2$IrO$_4$. Although its single t$_{2g}$ electron orders magnetically, Ba$_2$NaOsO$_6$ shows no evidence of an expected orbital order that causes Jahn-Teller distortion and should destroy its cubic symmetry. The origin of ferromagnetic order is unresolved, and evidently subtle since isostructural, isovalent, and also Mott insulating Ba$_2$LiOsO$_6$ orders antiferromagnetically. Specifically, the requirement of spin-orbit coupling to describe Ba$_2$NaOsO$_6$ makes it a Dirac-Mott insulator. In addition, recently reported x-ray magnetic circular dichroism (XMCD) for Ba$_2$NaOsO$_6$ has not been interpreted until now. Signals measured at the two osmium L edges are unusual in that one is zero (L$_3$), to a good approximation. Moreover, the edge with a null signal is not the one in previous measurements at Ir edges.

An interpretation of XMCD signals requires the magnetic space group. Ba$_2$NaOsO$_6$ forms an orthorhombic magnetic structure depicted in Fig. 1 below a transition temperature ≈ 7 K. Ferromagnetic layers are strongly canted with respect to the neighbouring layers, with Pnn′m′ the associated magnetic space group. Osmium ions (Os$^{7+}$, 5d$^1$) occupy Wykoff positions with low spatial symmetry. Together with a specified Kramers doublet, Pnn′m′ leads to satisfactory XMCD signals. Specifically, the L$_3$ signal can be zero from interference between a component in the ground state quartet (total angular momentum j = 3/2) and an excited doublet (j = 1/2). A calculated Os saturation magnetic moment accords with previous measurements. Calculations of the two mentioned properties reported here make extensive use of Racah algebra, and non-trivial XMCD reduced matrix elements listed in an Appendix.

Structural, chemical and magnetic properties of $Ba_2NaOsO_6$ along with $Ba_2MgReO_6$ ($Re^{6+}$) have been studied using various experimental techniques and theoretic methods [1-11]. The chemical structure of $Ba_2NaOsO_6$ at room temperature is $Fm\bar{3}m$ [2]. Da Cruz Pinha Barbosa *et al.* derive the cited long-range magnetic structure from neutron powder diffraction patterns (Fig. 1(d) [9]), and Lu *et al.* exploit NMR to the same end (Fig. 4(d) [7]). Synchrotron x-ray diffraction patterns reveal the chemical and magnetic structures of single crystals of $Ba_2MgReO_6$ (Fig. 1 (c) [8]). The osmium saturation magnetic moment is small [2], and it would be identically zero if the Kramers doublet was an unspoiled quartet (Eq. (8) [3]). In this case, the ground state is a quartet (total angular momentum j = 3/2) with an excited doublet (j = 1/2). An atomic model of a Kramers doublet can be useful in reaching a tenable understanding of magnetic properties when covalent bonding is strong [6, 11]. It is created from spin-orbital product states (spin = 1/2, orbital angular momentum = 2) with quantum numbers J, M (J = 3/2 and 5/2). Here, parameters account for Os 5d strongly hybridized with p orbitals that are close in energy [2].

## II. MATERIAL PROPERTIES

The magnetic space group of $Ba_2NaOsO_6$ is derived by a standard procedure using experimental results in Refs. [2, 7, 8, 9]. On the basis of the currently available data, it is beyond reasonable doubt that the ordered magnetic structures of $Ba_2NaOsO_6$ and $Ba_2MgReO_6$ are the same. Symmetry properties of the rhenate material have been thoroughly discussed with a view to future Bragg diffraction experiments [12]. We assert that, osmium ions occupy centrosymmetric sites in orthorhombic Pnn′m′ (No. 58.398, magnetic crystal class mm′m′ [13]), namely, Wyckoff positions 2a. A product of a dyad axis of rotation symmetry along the crystal c axis and time reversal symmetries 2′/m′ ($C_i$) restricts dipoles to the basal plane. According to $^{23}Na$ NMR spectra measured and analysed by Lu *et al.* there are two distinct magnetic sites Na(Os) [7]. The authors appeal to the canted magnetic structure in Fig. 1 for an interpretation of the spectra. This is not a viable interpretation, however, because Pnn′m′ possesses only one Na(Os) site.

## III. ELECTRONIC MULTIPOLES

Neumann's Principle places restrictions on Os multipoles $\langle T^K_Q \rangle$ of rank K and (2K + 1) projections in the interval $-K \leq Q \leq K$ used to formulate bulk properties, e.g., dichroic signals, and Bragg diffraction patterns [12, 14-16]. Angular brackets about the spherical tensor operator denote a time-average (expectation value). Site symmetry 2′/m′ for Wyckoff positions 2a includes inversion, and invariance of axial multipoles $\langle T^K_Q \rangle$ under the dyad operation $2_z′$. It is achieved by the identity $\sigma_\theta (-1)^Q = +1$, where the time signature $\sigma_\theta = +1$ (time-even, charge like) or $\sigma_\theta = -1$ (time-odd, magnetic). For $\sigma_\theta = -1$ projections Q are odd, and magnetic dipoles (K = 1) are confined to the basal plane. Our phase convention for real and imaginary parts labelled by single and double primes is $\langle T^K_Q \rangle = [\langle T^K_Q \rangle' + i\langle T^K_Q \rangle'']$. A tetragonal cell (ξ, η, ζ) used in [12] is shown in Fig. 1, with Cartesian dipoles $\langle T^1_\xi \rangle = -\sqrt{2} \langle T^1_{+1} \rangle'$ and $\langle T^1_\eta \rangle = -\sqrt{2} \langle T^1_{+1} \rangle''$.

XMCD signals are derived from an electronic structure factor $\Psi^K_Q$ reported in an Appendix and Ref. [12]. It is a sum of axial multipoles $\langle T^K_Q \rangle$ in a unit cell that exploits relations between Wyckoff positions [13]. For bulk properties of Os ions,

$$\Psi^K_Q(Pnn'm') = [\langle T^K_Q \rangle + (-1)^Q \langle T^K_{-Q} \rangle], \qquad (1)$$

in which site symmetry demands even (K + Q). A complex conjugate $\langle T^K_Q \rangle^* = (-1)^Q \langle T^K_{-Q} \rangle$, and $\Psi^K_Q(Pnn'm')$ in Eq. (1) is purely real. Bulk magnetism is determined by $\Psi^1_{+1}(Pnn'm') = 2 \langle T^1_{+1} \rangle' = -\sqrt{2} \langle T^1_\xi \rangle$.

## IV. MAGNETIC MOMENTS AND XMCD SIGNALS

Symmetry of Wykoff positions 2a in Pnn'm' requires odd Q for a magnetic multipole. In consequence, $[|u\rangle + f\,|\hat{u}\rangle]$ is a suitable wavefunction for an osmium ion, where $|u\rangle$ is a Kramers state, $|\hat{u}\rangle$ its conjugate, and the phase $\phi$ in $f = \exp(i\phi)$ is to be determined [12, 17]. The conjugate (time-reversed) state is constructed with the rule $c|j, m\rangle \rightarrow c^* \{(-1)^{j-m} |j, -m\rangle\}$, where c is a classical number. Using the identity $\langle u|T^K_Q|u\rangle = -\langle \hat{u}|T^K_Q|\hat{u}\rangle$, the corresponding expectation value,

$$\langle T^K_Q \rangle = (1/2)\,[f\,\langle u|T^K_Q|\hat{u}\rangle + f^*\,\langle \hat{u}|T^K_Q|u\rangle], \qquad (2)$$

with $f^* = \exp(-i\phi)$. Matrix elements in Eq. (2) are calculated with a standard prescription and known reduced matrix elements (RMEs) [16, 18]. Some aspects of the calculations are collected in an Appendix. Regarding the Os atomic wavefunction, spin-orbital product states $\{|\pm 1/2\rangle\,|l = 2, m\rangle\}$ are re-written in terms of angular momenta J = 3/2 and J = 5/2. The Kramers doublet taken to be,

$$|u\rangle = [|3/2, 1/2\rangle + t\,|5/2, -3/2\rangle]\,[1 + |t|^2]^{-1/2}, \qquad (3)$$

where t is a complex mixing parameter. Note that $\langle \hat{u}|T^K_Q|u\rangle$ does not include $t^*$, and $\langle u|T^K_Q|\hat{u}\rangle$ does not include t. Allowed projections in $\langle u|T^K_Q|\hat{u}\rangle$ are Q = ±1, −3, with Q = ±1, +3 for $\langle \hat{u}|T^K_Q|u\rangle$. A compatibility of calculated and measured properties of Os ions is our justification for Eq. (3).

A calculation of the magnetic moment (**L** + 2**S**) based on Eq. (3) is outlined in an Appendix. After extensive algebra the final result is,

$$\langle (\mathbf{L} + 2\mathbf{S}) \rangle_{+1} = (\sqrt{2}/5)\,[2f + t\,f^*\,\sqrt{3}]\,[1 + |t|^2]^{-1}, \qquad (4)$$

with Cartesian components $\langle (\mathbf{L} + 2\mathbf{S})_\xi \rangle = -\sqrt{2}\,\langle (\mathbf{L} + 2\mathbf{S})_{+1} \rangle'$ and $\langle (\mathbf{L} + 2\mathbf{S})_\eta \rangle = -\sqrt{2}\,\langle (\mathbf{L} + 2\mathbf{S})_{+1} \rangle''$. A discussion of the results is deferred to the next section.

Measured XMCD signals from $Ba_2NaCaOsO_6$ held a temperature of 3.5 K are shown in Fig 3d of Ref. [11]. Calculated XMCD signals use RMEs for $\mathbf{T}^K$ at $L_2$ and $L_3$ edges reported in an Appendix. They employ an axial electric dipole - electric dipole (E1-E1) absorption event [16]. The electronic structure factor Eq. (1) yields the following signals,

$$\Psi^1{}_{+1}(L_2) = -(1/9)\cos(\phi)\,[1 + |t|^2]^{-1}, \tag{5}$$

$$\Psi^1{}_{+1}(L_3) = \cos(\phi)\,[-2 + t\,9\sqrt{3}]\,(225\,[1 + |t|^2])^{-1}.$$

Note that orbital angular momentum $\langle \mathbf{L} \rangle$ obeys a sum rule [14, 16],

$$\langle \mathbf{T}^1(L_2) \rangle + \langle \mathbf{T}^1(L_3) \rangle = -\langle \mathbf{L} \rangle/(10\sqrt{2}), \tag{6}$$

which can be derived from Eq. (73) in Ref. [16].

## V. DISCUSSION AND SUMMARY

XMCD signals Eq. (5) exhibit properties captured in measurements at the L edges of osmium with a sample temperature = 3.5 K [11] (Fig. 3 and Section C in SM, with minimal information about powdered $Ba_2NaCaOsO_6$ used as a host material). According to the authors, signals measured at osmium $L_2$ and $L_3$ absorption edges are of one sign, and the $L_3$ signal is relatively small; see the text immediately below Fig. 3 in Ref. [11]. According to Eq. (6), XMCD signals of one sign are consistent with a significant orbital moment on the resonant ion. Evidently, the extreme result $\Psi^1{}_{+1}(L_3) = 0$ is achieved for a real mixing parameter $t = (2/9\sqrt{3})$ in the Kramers doublet Eq. (3), which then implies that a weak $L_3$ signal is due to interference of atomic states $J = 3/2$ and $J = 5/2$. The sign of $\Psi^1{}_{+1}(L_2)$ is fixed by a single parameter, namely, $\cos(\phi)$. Interestingly, null values of an $L_2$ signal are known in other materials, including $Sr_2IrO_4$ [17].

A plausible interpretation of the magnetic moment Eq. (4) uses $t = (2/9\sqrt{3})$ that leaves XMCD $\propto \Psi^1{}_{+1}(L_3) = 0$. Adding the constraint $\langle (\mathbf{L} + 2\mathbf{S}) \rangle_\xi = \langle (\mathbf{L} + 2\mathbf{S}) \rangle_\eta$ yields $\langle (\mathbf{L} + 2\mathbf{S}) \rangle_\xi = 0.546$, which matches the moment measured in the paramagnetic phase [1, 2]. A sample of $Ba_2NaOsO_6$ held at about half of the magnetic ordering temperature (Fig 3d in Ref. [11]) possesses a ferromagnetic magnetic moment $\approx 0.2$ [1, 2, 4]. One achieves $\langle (\mathbf{L} + 2\mathbf{S}) \rangle_\xi = 0.2$ and $\langle (\mathbf{L} + 2\mathbf{S}) \rangle_\eta = 0$ using $t = -0.717$ and $\phi = 0$ in Eq. (4). Corresponding XMCD signals accord with measured signals [11], being of one sign and in a ratio $\{\Psi^1{}_{+1}(L_2)/\Psi^1{}_{+1}(L_3)\} = 1.90$. The mixing parameter, $t$, must absorb effects of shielding by electron back-transfer between osmium and neighbouring oxygen ions (covalency). Indeed, Gangopadhyay and Pickett [6] claim that the net osmium moment lies primarily on the oxygen ions.

Summarizing, we assign the long-range magnetic order of $Ba_2NaOsO_6$ depicted in Fig. 1 to the orthorhombic space group Pnn′m′ (No. 58.398) [12, 13]. Future measurements of magnetic Bragg diffraction patterns can be confronted with published diffraction patterns for Pnn′m′ [12]. The mentioned study includes allowed structural distortions. Down to a sample temperature of 10 K there is no evidence of distortions away from cubic symmetry (neutron powder diffraction patterns [9]). Symmetry of the Wyckoff positions imposes restrictions on osmium electronic multipoles, and the magnetic structure leads to Eq. (1) for the XMCD signal. It is an exact result, likewise all our work leading to results in Eqs. (4) and (5) [16, 18]. The brevity of Eq. (3) for the Kramers state is commendable.

**Acknowledgements** Correspondence with Dr S. Agrestini, Dr D. D. Khalyavin, and Professor G. van der Laan.

## APPENDIX

A matrix element in Eq. (2), for example, is calculated with [19],

$$\langle JM|T^K_Q|J'M'\rangle = (-1)^{J-M} \begin{pmatrix} J & K & J' \\ -M & Q & M' \end{pmatrix} (J\|T^K\|J'), \quad (A1)$$

where the right-hand side includes a product of a 3j-symbol (or a Clebsch-Gordan coefficient) and a reduced matrix element (RME) = $(J\|T^K\|J')$. XMCD signals from an E1-E1 absorption event depend on dipole matrix elements, namely, $(3/2\|T^1\|3/2) = -(1/15)\sqrt{(2/15)}$, $(3/2\|T^1\|5/2) = -(5/2\|T^1\|3/2) = -(1/5)\sqrt{(3/10)}$ for the $L_3$ edge, and $(3/2\|T^1\|3/2) = -(1/3)\sqrt{(5/6)}$ for the $L_2$ edge. The RMEs are derived from Eq. (73) in Ref. [16]. Expectation values of $\mathbf{T}^1$ follow from Eq. (2), and for K = 1 coefficients of f and f* are, respectively,

$$\langle u|T^1_{+1}|\hat{u}\rangle = \sqrt{(2/15)}\,(3/2\|T^1\|3/2)\,[1 + |t|^2]^{-1},$$

$$\langle \hat{u}|T^1_{+1}|u\rangle = -\sqrt{(2/5)}\,t\,(3/2\|T^1\|5/2)\,[1 + |t|^2]^{-1}. \quad (A2)$$

The expectation value of the magnetic moment also follows from Eqs. (2) and (A2).

The coefficient of f in Eq. (2) is seen to be diagonal in J. For the magnetic moment it can be calculated with $(\mathbf{L} + 2\mathbf{S}) = g\mathbf{J}$ using an RME = $[J(J+1)(2J+1)]^{1/2}$ and a Landé factor g = 4/5 (S = 1/2, L = 2). Conversely, the coefficient of f* in Eq. (2) is an off-diagonal matrix element, namely, that of $\mathbf{S}$ in $(\mathbf{L} + 2\mathbf{S}) = (\mathbf{J} + \mathbf{S})$ between states J = 3/2 and 5/2. The required RME is obtained from Eq. (7.1.7) in Ref. [19].

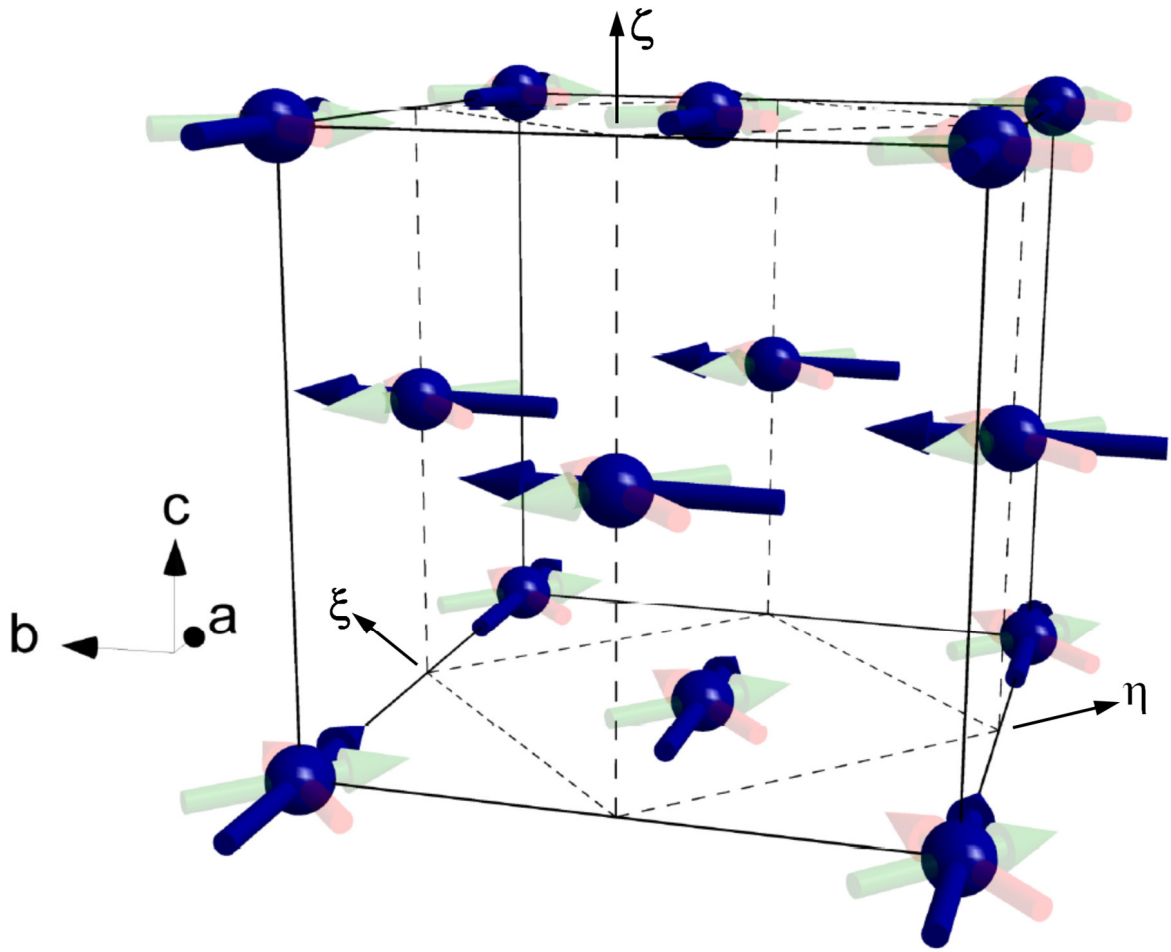

**FIG. 1**. Ferro- and antiferromagnetic osmium dipole components of the ordered magnetic structure $Ba_2NaOsO_6$ are depicted by transparent red and green arrows, respectively. A tetragonal basis labelled (ξ, η, ζ) is introduced in Section 3 [12].